\documentclass[pra,showpacs,preprintnumbers,amsmath,amssymb,tightenlines,twocolumn,superscriptaddress]{revtex4}

\usepackage{graphicx}
\usepackage{dcolumn}
\usepackage{bm}
\usepackage{epsfig}
\usepackage{stmaryrd}
\usepackage{color}
\usepackage{psfrag}

\newcommand{\bra}[1]{\langle\,{#1}\, |}
\newcommand{\ket}[1]{|\,{#1}\,\rangle}
\newcommand{\braket}[2]{\mbox{$\langle\,{#1}\, | \,{#2}\,\rangle$}}

\newcommand{\vek}[1]{\boldsymbol{#1}}
\newcommand{\cref}[1]{chapter~\ref{#1}}

\newcommand{\Cref}[1]{Chapter~\ref{#1}}

\begin{document}

\title{Quantum Dynamics Simulation with Classical Oscillators}

\author{John~S.~Briggs }
\email{briggs@physik.uni-freiburg.de}
\affiliation{Max Planck Institute for the Physics of Complex Systems,
N\"othnitzer Strasse 38, 01187 Dresden, Germany}
\author{ Alexander Eisfeld}
\email{eisfeld@pks.mpg.de}
\affiliation{Max Planck Institute for the Physics of Complex Systems,
N\"othnitzer Strasse 38, 01187 Dresden, Germany}

\begin{abstract}
In a previous paper [J.~S.~Briggs and A.~Eisfeld, Phys.~Rev.~A {\bf{85}} 052111 (2012)] we showed that the time-development of the complex amplitudes of
$N$ coupled quantum states can be mapped by the time development of positions and velocities of $N$ coupled classical oscillators. Here we examine to 
what extent this mapping can be realised to simulate the "quantum" properties of entanglement and qubit manipulation. By working through specific examples, e.g. of quantum gate operation, we seek to illuminate quantum/classical differences which hitherto have been treated more mathematically. 
In addition we show that important 
quantum coupled phenomena, such as the Landau-Zener transition and the occurrence of Fano resonances can be simulated by classical oscillators.
\end{abstract}
\maketitle
\section{Introduction}
Scattered throughout the recent literature, responding to a renewed interest in studying the ostensibly unique properties of quantum systems, there have been many papers devoted to demonstrating that some aspects of quantum dynamics can be reproduced by classical systems. These classical systems are often assemblies of classical oscillators and the equivalence to quantum coupled systems stems essentially from the mathematical correspondence between classical eigenfrequencies and quantum eigenvalues.
Depending on the system, sometimes the correspondence presented has been merely an analogy, sometimes it has been more exact.
 However, very few papers point out that the mapping of quantum dynamics as represented by the time-dependent Schr\"odinger equation (TDSE), can be traced right back to the very first paper applying this equation by Dirac \cite{Dirac_PRSA_114}, who showed that the first-order time-dependent coupled equations for quantum state amplitudes are identical to classical Hamilton equations. Much later this equivalence was discovered independently by Strocchi \cite{St66_36_} but without application.

 In previous publications \cite{BrEi11_051911_,EiBr12_046118_,BrEi12_052111_} we have extended this analysis and in particular shown how, for Hermitian quantum Hamiltonians, the quantum dynamics corresponds  specifically to the classical mechanics of the generalised coupled motion of mechanical or electrical oscillators. 
In particular we showed that, although an exact mapping of the TDSE is possible, it can lead to rather complicated to realize classical coupled equations involving simultaneously position and momentum coupling of the oscillators. More standard classical equations, 
where the coupling between the oscillators is only via the position coordinates, can be achieved in a weak-coupling approximation which we referred to as the "realistic coupling approximation" (RCA). Indeed almost all previous publications simulating quantum dynamics with classical coupled systems implicitly assume the RCA and write down classical oscillator equations without reference to the exact Dirac mapping. One aim of our previous work was to assess the accuracy of the RCA. 
To this end, as example, we have shown  \cite{BrEi11_051911_,EiBr12_046118_} that the coherent transfer of electronic excitation between coupled molecules (for example in the photosynthetic unit), often ascribed as due to a manifestation of quantum entanglement, can be simulated exactly by transfer between coupled classical harmonically-oscillating electric dipoles  and the RCA can give an excellent approximation to the exact dynamics.

In this paper we wish to explore the consequences of the Dirac mapping further, firstly  by considering to what extent quantum aspects such as entanglement and quantum gate operation can be reproduced by purely classical motion. As in our previous studies, these applications involve mapping the quantum dynamics of real, Hermitian time-independent  Hamiltonians.  We will show that certain aspects of entanglement measures and quantum gate operation are readily simulated classically. In the more general case of time-dependent or non-Hermitian Hamiltonians we demonstrate that quantum interference effects and non-adiabatic transitions can be simulated also.

 However, it is also illustrative to see where differences between the quantum and the classical case arise. 
In this way we hope to shed light on the oft-discussed problem of quantum/classical correspondence by discussing concrete examples. It will emerge that the key point of the Dirac mapping is that each and every quantum state must be assigned to a separate classical oscillator. Hence, if we have a given number of  "particles" (atoms, molecules, spin systems) with one level each, this system may be simulated by $N$ oscillators.  However, if each particle has many quantum states the total number of many-body states proliferates and correspondingly the number of classical oscillators required for the simulation. As the simplest but important example, consider a quantum system of $N$ two-level "particles" (qubits).  Each qubit can be represented by $2$  classical oscillators giving a total of $2N$ oscillators. However the $N$ qubits give rise to $2^N$ quantum states, say corresponding to increasing number of qubits in the upper state, all the way from zero to $N$. Hence the simulation of the general coupled system of qubits requires not $2N$ but $2^N$ classical oscillators. That this proliferation of classical systems with respect to their quantum counterpart, rather than entanglement \emph{per se} is what decides the possible exponential advantage of quantum compared to classical computing in performing certain algorithms has been emphasised before, particularly in Ref.~\cite{JoLi03_2011_}.

The modelling of quantum systems by classical oscillators goes right back to pre-quantum days when, for example, Lorentz \cite{Lo06_591_} and Holtsmark \cite{Ho25_722_} described absorption of light by atoms as the excitation of classical oscillating electric dipoles. This tradition was  extended into the quantum mechanics era in the work of Fano \cite{Fa60_451_}  on co-operative quantum states modelled by the eigenmodes of coupled classical dipoles and more recently by the simulation of Fano resonances by coupled oscillators \cite{JoSaKi06_259_}  or the analogous treatment of electromagnetic-induced transparency (EIT)  \cite{TaZhKo09_053901_}. Many more examples have been given of classical oscillator simulation of various few-level quantum systems and several of these papers are referred to in Ref.~\cite{BrEi12_052111_}.
Hence, as a second problem, we examine the mapping of the quantum dynamics of more general time-dependent and non-hermitian Hamiltonians. Again it is instructive to consider specific examples and we have chosen two systems ubiquitous in quantum physics, namely, the Landau-Zener avoided crossing of two quantum states and the interaction of a discrete quantum state with a continuum of states, giving rise to so-called "Fano" resonances. In both problems we examine the exact mapping and the utility of the simpler RCA coupled equations.

Recently, Skinner \cite{Sk13_012110_}  has extended our previous work \cite{BrEi11_051911_,EiBr12_046118_,BrEi12_052111_} on Dirac mapping to include complex hermitian Hamiltonians and more general dissipative systems. In particular,  the interesting suggestion is made that in these systems a simulation is more easily realised by doubling the number of  classical oscillators. This suggestion is discussed in more detail below.

 The plan of the paper is as follows. In section II  we introduce the basic classical equations which map the time-development of quantum systems whose wavefunction is expanded in some basis set. In section III we examine the question of entanglement measures and show simply how some measures have exact classical counterparts.
Also we show that certain operations of quantum computing can be performed readily by classical oscillators. 
 Our aim here is to give concrete, realisable classical systems capable of demonstrating this correspondence in the laboratory. 
It emerges that an arrangement of  $N$ coupled oscillators can mimic \emph{exactly} a quantum system of $N$ coupled two-level systems, so long as the excitation is confined to one quantum. However, in the more general case, including all states with up to $N$ quanta giving a total of $2^N$ quantum states, then $2^N$ oscillators are required to achieve a simulation.

  In section IV we discuss time-dependent Hamiltonians and examine simulation of  the celebrated Landau-Zener non-adiabatic transition as example. Section V is devoted to the question of non-hermitian Hamiltonians in the Schr\"odinger equation. The consequences of the results are discussed in the concluding section VI.

\section{ The Quantum and Classical Equivalent Equations}

In ref.~\cite{BrEi12_052111_}, to be referred to as paper I, it was shown how the coupled time-dependent Schr\"odinger equation for the complex amplitudes of a quantum level system involving a finite number of levels can be mapped to
the Newton equations of the same number of coupled classical oscillators. Here we re-iterate this mapping briefly for the case of a real and time-independent quantum Hamiltonian. Later we will extend to complex and time-dependent Hamiltonians. 

The basis set expansion,
\begin{equation}
\label{eq:basis_expansion}
\ket{\Psi(t)} = \sum_n c_n(t) \ket{\phi_n},
\end{equation}
of solutions of the time-dependent Schr\"odinger equation (TDSE), where the $c_n$ are complex co-efficients and $\ket{\phi_n}$ denotes an arbitrary orthonormal time-independent basis, leads to a set of coupled equations (we use units such that $\hbar = 1$),
\begin{equation}
\label{eq:TDSE}
i\dot c_n(t) = \sum_m H_{nm} c_m(t).
\end{equation}

 In the special case that all matrix elements $H_{nm}$ are real the TDSE coupled equations (\ref{eq:TDSE}) are equivalent to classical Hamilton equations i.e. $c_n = z_n$ with $z_n \equiv (q_n + ip_n)/\sqrt{2}$, and $p_n$ and $q_n$ real momenta and positions, if the 'classical' Hamiltonian function is taken as the expectation value of the quantum Hamiltonian i.e.
\begin{equation}
\mathcal{H} = \bra{\Psi(t)}H\ket{\Psi(t)} = \sum_{nm}c^*_n(t) H_{nm} c_m(t).
\end{equation}
 the Hamiltonian becomes
\begin{equation}
\label{eq:classHam}
\mathcal{H} = \frac{1}{2} \sum_{nm} H_{nm}( q_nq_m + p_np_m),
\end{equation}
which is that of coupled real harmonic oscillators. Note that the coupling is of a very special form in which there is both bi-linear position and momentum off-diagonal coupling with exactly the same coupling strengths. We will call the equations of motion derived from this Hamiltonian the "p-and q-coupled equations".

The Hamilton equations resulting from a time-independent quantum Hamiltonian are
\begin{equation} 
\label{eq:Hameqns}
\dot q_n = \sum_{m} H_{nm}p_m,\qquad \dot p_n = -\sum_{m} H_{nm}q_m.
\end{equation}
Taking the time derivative of the $\dot q_n$ equation and inserting the $ \dot p_n$ equation one obtains 
\begin{equation}
\ddot q_n= -\sum_{mm'}H_{nm}H_{mm'}q_{m'}.
\end{equation}
A similar equation can be obtained for the momenta $p_n$.

Symbolically, writing $q$ and $p$ as vectors and $H$ as a matrix the above equations are
\begin{equation}
\label{eq:hameq}
\mathbf{\dot q} = \mathbf{H} \mathbf{p}\qquad \mathbf{\dot p} = -\mathbf{H} \mathbf{q}
\end{equation}
and formally
\begin{equation}
\mathbf{\ddot q} = \mathbf{H \dot{p}} = -\mathbf{H}^2\mathbf{q}
\end{equation}
which are a set of coupled oscillator equations and can be solved for $\mathbf{q}(t)$ and  $\mathbf{\dot q}(t)$. 
Typically, in a physical realisation of the coupled classical oscillators one would measure the positions $q_n$ and the velocities $\dot{q}_n$. However, to construct the quantum amplitudes one needs $q+ip$, i.e.~the momenta $p_n$.  
The momenta at time $t$ can be calculated from
 \begin{equation}
\label{eq:pdef}
\mathbf{p} = \mathbf{H}^{-1} \mathbf{\dot q}.
\end{equation}
The set of  complex amplitudes, the vector $\mathbf{z}$, is constructed as
\begin{equation}
\label{eq:zclass}
 \mathbf{z} = \frac{1}{\sqrt2} (\mathbf{q} + i\mathbf{p}).
\end{equation}
From the Hamilton Eqs.~(\ref{eq:hameq}) we have
\begin{equation}
\label{eq:ddotz}
\mathbf{\ddot z}  + \mathbf{H}^2\mathbf{z} = 0
\end{equation}
Similarly the Schr\"odinger equation (\ref{eq:TDSE}) is written
\begin{equation}
\label{eq:Shcoupled}
i\mathbf{\dot{c}} = \mathbf{H}\mathbf{c}
\end{equation}
or
\begin{equation}
\label{eq:Clcoupled}
\mathbf{\ddot c} + \mathbf{H}^2\mathbf{c} = 0
\end{equation}
which is exactly the classical equation (\ref{eq:ddotz}) and makes the similarity to the equations of a set of coupled classical oscillators obvious, at least formally. Hence the p- and q-coupled classical equations and the coupled quantum Schr\"odinger equations are identical.

It is clear that the form of the above equations (\ref{eq:Clcoupled}) and (\ref{eq:ddotz})  are also applicable when $\mathbf{H}$ is complex. This will be illustrated in section \ref{sec:dissipative}
where coupling of dissipative states is considered.

\section{Entanglement and Quantum Gates}
\subsection{Entanglement and J Aggregates.}
One prominent example of an entangled quantum system is the excitonic J band formed by certain aggregates of dye molecules. The eigenstates (known as Frenkel excitons)  forming the J-band states of such a system are linear 
combinations of states in which one molecule is electronically excited and all others in their electronic ground state. We call these states $\ket{\pi_n}$ where the $n'th$ molecule is excited.
Note that within this so-called one-exciton manifold the states $\ket{\pi_n}$ can be readily identified with the states $\ket{\phi_n}$ of Eq.~(\ref{eq:basis_expansion}). Therefore, quantum mechanical excitation of a certain molecule can be associated, within the classical mapping, with oscillation of the corresponding classical oscillator \cite{BrEi11_051911_,EiBr12_046118_,BrEi12_052111_}.

 Thilagam \cite{Th11_135306_} has studied the entanglement dynamics of a J aggregate by calculating the entanglement measures of von Neumann entropy and concurrence.  She  suggests that "the entangled properties highlight the potential in utilizing opto-electronics properties of J-aggregate systems for quantum information processing".
These measures are constructed from the density matrix whose elements are composed of bi-linear products of the complex amplitudes $c_n$. However, since these complex numbers are identical to the classical $z_n$ numbers, it is clear that these entanglement measures are reproduced \emph{exactly} by the classical dynamics. Hence this is a completely classical reproduction of "quantum entanglement".

We emphasise that the simulation of the entanglement of quantum {\it states}  by classical oscillators is quite general  as far as the entanglement measures (e.g.\ entropy, concurrence, negativity, quantum discord) are calculated using density matrix elements, written as binary products of amplitudes, since these are identical in quantum and classical dynamics. Note however, in the J-aggregate excited state considered by Thilagam, we only have one excitation shared between $N$ oscillators and hence our excited-state Hilbert space is of dimension $N$ and can be mapped exactly onto $N$ classical oscillators where initially only one oscillator is excited. 
 Thus it is the restriction to the one-exciton space that allows one to associate a localised excitation on a certain monomer with a single classical oscillator. In the case of more than one excitation in the system, there will be quantum states which contain excitation on two (or more) molecules.  Such a state would map to its own oscillator, thus the simple correspondence molecule-oscillator no longer holds. This is another example of the necessity, mentioned in the Introduction, to have more classical oscillators as the number of excited quantum two-level systems increases.

In view of the ability of classical coherent coupled motion to reproduce certain aspects of what is viewed as a purely quantum effect, it appears apposite to examine other features of quantum information processing to ascertain which aspects are reproduced by classical coupled oscillator motion. To do this the basic ingredients of quantum information must be simulated classically. The most important of these are definition of a qubit, its rotation on the Bloch sphere and the coupling (entanglement) of two qubits in the construction of quantum gates. These questions are examined in the following sections.

\subsection{The classical qubit}
We first define a classical state of a pair of oscillators which corresponds to the state of a single quantum 2-level system, a qubit. Then we show that by changing the amplitude and phase of the classical oscillators we can perform arbitrary rotations on the Bloch sphere.
In paper I we have pointed out that the complex amplitudes of monomer eigenstates coupled into a quantum dimer
are identical to suitably-defined classical complex amplitudes of coupled oscillators. Essentially these results are the same as given in section IIIA but restricting the excitation to $N=2$ monomers. Here the states of  coupled classical oscillators will be defined  in a way that allows quantum gate operations to be performed with them. 
  
\subsubsection{The single qubit:  quantum monomer and two coupled oscillators}
We consider the quantum monomer to consist of only two states, a ground state and an excited state denoted by $\ket{0}$ and $\ket{1}$ respectively. These two states can constitute a qubit.

To define a qubit, in the standard notation we  consider a linear superposition of the two states, $\ket {0}$ and $\ket {1}$ with complex coefficients. Generally, up to an irrelevant overall phase, one can take the co-efficient of $\ket{0}$ to be real and non-negative and define a state on the Bloch sphere  by
 \begin{equation}
 \label{eq:Bloch state}
\ket{\psi} = \cos (\theta/2) \ket{0} + e^{i\phi} \sin(\theta/2) \ket{1},
\end{equation}
where the parameters $\theta$ and $\phi$ are the usual angles defining the unit sphere.
Then the state is normalised $\braket{\psi}{ \psi} = 1$, the state $\ket{0}$ is at the north pole and $\ket{1}$ at the south pole. In addition particular superpositions of these  states can be used as basis.
 Clearly the position on the Bloch sphere is defined by a single complex number subject to the normalisation condition. According to the mapping prescription, a classical oscillator corresponds to each of the two states. The motion of each oscillator is uniquely defined by  the two real numbers of maximum amplitude and phase, from which the complex number $q+ip$ may be specified. Hence there are four real numbers specifying the classical "qubit". However, as in the quantum case only relative phase is significant, which eliminates one number. In the classical case, conservation of total energy of the two oscillators gives a relation between amplitudes and plays the role of normalisation in the quantum case. Hence only two real numbers specify the complex number locating the pair of oscillators on the Bloch sphere and the state of the oscillators can also be described symbolically by Eq.~(\ref{eq:Bloch state}). \\
In the quantum case the rotation on the Bloch sphere is achieved by allowing the states $\ket{0}$ and $\ket{1}$ to interact for a time to form a new linear superposition. This time development is a unitary transformation.  The same is true classically, the corresponding two oscillators interact for a time and unitrarity is mapped to the classical dynamics by ensuring energy conservation during the interaction. Since this interaction of two states is the cornerstone of our simulation we examine the transformation in some detail.\\
We consider two arbitrary quantum states, for simplicity but without great loss of generality which we take to be degenerate in energy (in real systems slight non-degeneracy is often desirable to inhibit interaction but may be lifted by application of external fields). We will call the two states, as above for a qubit, $\ket{0}$ and $\ket{1}$.

A coherent superposition of  $\ket{0}$ and $\ket{1}$ is achieved by  switching on an interaction $V$ between the two states for a certain time. The coupled qubit has $+$ and $-$ eigenstates of the form,
\begin{equation}
\label{eq:dimereigen}
\ket{\psi_{\pm}} = \frac{1}{\sqrt 2}(\ket{0} \pm \ket{1})
\end{equation}
with eigenenergies $\epsilon_{\pm} = \epsilon \pm V$, where $\epsilon$ is the energy of the pair of states (which can be put to zero). 
Then one can show (see paper 1) that a general solution of the time-dependent Schr\"odinger Equation  (TDSE) can be written,
\noindent
\begin{equation}
\begin{split}
\label{eq:c1quant}
c_1(t) =  \exp[-(i/\hbar) \epsilon t]  \cos [Vt/\hbar]\\
c_2(t) = - i \exp[-(i/\hbar) \epsilon t]  \sin [Vt/\hbar]
\end{split}
\end{equation}
which are the exact quantum solutions and describe a periodic transfer of energy between the two states. A particular change in amplitude and phase can be achieved by choosing the coupling time and so a rotation on the Bloch sphere is performed.

Exactly the same transformation can be made using the two classical oscillators.
 When mapped to the Hamilton equations, the Hamiltonian of the two coupled quantum states gives rise to classical equations of motion for the displacements $q_1$ and $ q_2$ of two identical coupled pendula of natural frequency $\omega$. The  coupled oscillator equations (derived in paper I) are,
\begin{equation}
\begin{split}
\ddot q_1 + (\omega^2 + V^2)q_1& = -2\omega Vq_2\\
\ddot q_2 + (\omega^2 + V^2)q_2 &= -2\omega Vq_1
\end{split}
\end{equation}
In the usual way these symmetric equations can be diagonalised by the transformation $q_{\pm} = (q_1 \pm q_2)/\sqrt 2$ to give normal modes satisfying the uncoupled equations
\begin{equation}
\ddot q_{\pm} + (\omega \pm V)^2q_{\pm} = 0
\end{equation}
with eigenfrequencies $\Omega_{\pm} = \omega \pm V$. As they should be, these are exactly the eigenfrequencies $\epsilon \pm V$ of the quantum two-state problem derived above. Then one can show that the classical complex amplitudes $z_1$ and $z_2$ obey exactly the same equations as the quantum amplitudes 
$ c_1, c_2$ of Eq.(\ref{eq:c1quant}). Again by choosing interaction time and strength of coupling the relative amplitudes can be changed arbitrarily. A relative phase change simply requires a change of the phase of one oscillator.
Accordingly one sees that the operation leading to the quantum mixing of the two qubit states, or rotation on the Bloch sphere, also can be performed exactly by a pair of classical coupled oscillators.\\
In particular the Hadamard gate is defined
 \begin{equation}
 \mathbb{H} = \frac{1}{\sqrt{2}} \left[ \begin{array}{ c c }
1 & 1 \\	
1& -1
 \end{array} \right]
 \end{equation} 
and transforms the basis $\ket{0}$ and $\ket{1}$ into the mixed basis $\ket{\psi_+}$ and $\ket{\psi_-}$, i.e.
 \begin{equation}
\mathbb{H} \ket{0} = \frac{1}{\sqrt 2} (\ket{0} + \ket{1}) \equiv \ket{\psi_+}
\end{equation}
and
 \begin{equation}
\mathbb{H} \ket{1} = \frac{1}{\sqrt 2} (\ket{0} - \ket{1}) \equiv \ket{\psi_-}.
\end{equation}
Hence this operation produces eigenvectors of the interaction (which are actually eigenvectors of the Pauli matrix $\sigma_x$) from the two-state basis. Clearly this simple quantum gate can be simulated by the classical oscillators by bringing them into interaction to form the eigenmodes $q_\pm$.\\

 \subsubsection{Two qubits: quantum dimer and four coupled oscillators.}
 
 In the dimer composed of two qubits we will denote the states with a double index, the first referring to monomer $a$, the second to monomer $b$. Then the absolute ground state is denoted $ \ket{0}_a\ket{0}_b \equiv\ket{00}$.  The doubly-excited state is then $\ket{1}_a\ket{1}_b \equiv\ket{11}$. The two singly-excited states
 are  $\ket{0}_a\ket{1}_b \equiv\ket{01}$  and $\ket{1}_a\ket{0}_b \equiv\ket{10}$. The total of $4~ (= 2^2)$ non-interacting states 
 are designated
  \begin{equation}
  \begin{split}
&\ket{00} = 
   \left[ \begin{array}{ c  }
 1 \\	
 0 \\
 0 \\
 0 \\
 \end{array} \right],
   \ket{01} =
   \left[ \begin{array}{ c  }
 0 \\	
 1 \\
 0 \\
 0 \\
 \end{array} \right],\\
&  \ket{10} =
   \left[ \begin{array}{ c  }
 0 \\	
 0 \\
 1 \\
 0 \\
 \end{array} \right],
  \ \ket{11} =
   \left[ \begin{array}{ c  }
 0 \\	
 0 \\
 0 \\
 1 \\
 \end{array} \right].
 \end{split}
 \end{equation}
 In the notation of section II we make the identification $\ket{\pi_1} \equiv \ket{10} $ and $\ket{\pi_2} \equiv \ket{01} $.
To operate, the two qbits must be entangled through interaction, which must be on/off switchable.  
 The general entangled state is usually written in the non-interacting (computational) basis as
\begin{equation}
\ket{\psi} = \alpha \ket{00} + \beta \ket{01} + \gamma \ket{10} + \delta \ket{11}.
\end{equation}\\
Now we consider simulation of two-qubit gates.

The first, called the $\mathrm{SWAP}$ gate, involves mixing of only two of the four states which we take here to be the $\ket{10}$ and $\ket{01}$ states. This gate simply swaps the amplitudes of the two states involved and hence can be achieved as is done in the rotation of a single qubit. For example, starting with unit amplitude $c_1$ for state $\ket{10}$ and
zero amplitude $c_2$ for state $\ket{01}$, after an interaction time $t$ the amplitudes are given exactly by Eq.~(\ref{eq:c1quant}). The 
$\mathrm{SWAP}$ gate corresponds  to switching on interaction for a time $t= \pi/(2V)$  which transforms $\ket{10}$ into $-i\ket{01}$, i.e. 
  \begin{equation}
\mathrm{SWAP}  = 
\begin{bmatrix}
0&1\\
1&0
\end{bmatrix}.
\end{equation}

 More importantly, for a time $t= \pi/(4V)$  the entanglement corresponds to the $\mathrm{SQiSW}$ gate represented in the two-state space as
 \begin{equation}
\mathrm{SQiSW}  = \frac{1}{\sqrt 2}
\begin{bmatrix}
1&-i\\
-i&1
\end{bmatrix}
\end{equation}
 Beginning in the separable state $\ket{10}$ this gate produces the entangled state $(-i\ket{01} + \ket{10})/\sqrt 2$.
 This  $\mathrm{SQiSW}$ gate, which can be simulated by a pair of coupled oscillators, was employed for example in the quantum tomography experiment of Ref.~\cite{BiAnHo10_409_}.

The most important two-qubit quantum gate is the CNOT gate
which operates on the entangled wavefunctions of interest for quantum computing. The operation uses qbit $a$ as control bit and qbit $b$ as target bit. 
 The gate corresponds simply to the instruction : if $a$ is in the ground state, do not change $b$, if $a$ is in the excited state, then change the state of $b$, i.e.
 \begin{equation}
 \label{eq:CNOT}
 \begin{split}
&\mathrm{CNOT}\  \ket{00}  \rightarrow \ket{00}\\
&\mathrm{CNOT}\ \ket{01} \rightarrow \ket{01}\\
&\mathrm{CNOT}\ \ket{10} \rightarrow \ket{11}\\
&\mathrm{CNOT}\ \ket{11} \rightarrow \ket{10}\\
\end{split}
\end{equation}
 In the 4-dimensional computational 
basis  CNOT has the representation,
 \begin{equation}
 \mathrm{CNOT} = 
\begin{bmatrix}
1&0&0&0\\
0&1&0&0\\
0&0&0&1\\
0&0&1&0
\end{bmatrix}
\end{equation}
The CNOT gate can be decomposed into a sequence of operations involving rotations of the individual qubits and operation of the SQiSW gate which generates an entanglement of the degenerate
$\ket{01}$ and $\ket{10}$ states only.
The decomposition is
\begin{equation}
\begin{split}
{\rm CNOT}=R_y^a(-\pi/2)[ R_x^a(\pi/2) \otimes R_x^b(-\pi/2)] &\\\mathrm{SQiSW} R_x^a(\pi) \mathrm{SQiSW} R_y^a(\pi/2)
\end{split}
\end{equation}
In order to compare with operations on classical oscillators, we show in Appendix A how to follow this sequence of transformations through for a given initial state.

Now we consider the same sequence of operations performed with four identical classical oscillators. We assign a separate classical oscillator to each of the four quantum states.
We consider the quantum gate transformations sequentially and use matrix notation to indicate the couplings operating in each step. 
Consider first a rotation $R_y^a(\pi/2)\otimes \bf{1}_b$ operating on $\ket{00}$ as initial state. The result is
\begin{equation}
 \frac{1}{\sqrt 2}
\begin{bmatrix}
1&0&-1&0\\
0&1&0&-1\\
1&0&1&0\\
0&1&0&1
\end{bmatrix}
\left(\begin{array}{c}
1\\
0\\
0\\
0\\
\end{array}\right)
= \frac{1}{ \sqrt 2} \left(\begin{array}{c}
1\\
0\\
1\\
0\\
\end{array}\right)
\end{equation}
which corresponds exactly to $\ket{\Psi_{ab}}_2$ of Eq.~(\ref{eq:Psiab2}). The matrix represents the unitary transformation $\ket{00} \to (\ket{00} + \ket{10})/\sqrt 2$ i.e. to rotating qubit $a$ and can be achieved by coupling the two oscillators representing the two states of qubit $a$ only.\\
Similarly the $\mathrm{SQiSW}$ gate can be reproduced by coupling the oscillator $\ket{10}$, which is now in motion, to $\ket{01}$ for the appropriate time. This is the operation
\begin{equation}
 \frac{1}{\sqrt 2}
\begin{bmatrix}
1&0&0&0\\
0&\frac{1}{\sqrt 2}&\frac{-i}{\sqrt 2}&-0\\
0&\frac{-i}{\sqrt 2}&\frac{1}{\sqrt 2}&0\\
0&0&0&1
\end{bmatrix}
\left(\begin{array}{c}
1\\
0\\
1\\
0\\
\end{array}\right)
= \frac{1}{ \sqrt 2} \left(\begin{array}{c}
1\\
\frac{-i}{\sqrt 2}\\
\frac{1}{\sqrt 2}\\
0\\
\end{array}\right)
\end{equation}
again which corresponds exactly to the entangled state $\ket{\Psi_{ab}}_3$ of Eq.~(\ref{eq:Psiab3}). The correlated motion of the four oscillators now corresponds 
to the entangled state.\\
The next step is a rotation of this state by $R_x^a(\pi) \otimes \bf{1}_b$ given by
\begin{equation}
\begin{bmatrix}
0&0&-i&0\\
0&0&0&-i\\
-i&0&0&0\\
0&-i&0&0
\end{bmatrix}
\left(\begin{array}{c}
 \frac{1}{\sqrt 2}\\
\frac{-i}{\sqrt 2}\\
\frac{1}{2}\\
0\\
\end{array}\right)
= \frac{1}{ \sqrt 2} \left(\begin{array}{c}
\frac{-i}{\sqrt 2}\\
0\\
-i\\
-\frac{1}{\sqrt 2}\\
\end{array}\right)
\end{equation}
again which corresponds exactly to the entangled state $\ket{\Psi_{ab}}_4$ of Eq.~(\ref{eq:Psiab4}).
Although only a rotation of qubit $a$, since we have an entangled state this involves a change in the amplitude of all four oscillators and so looks to involve coupling all four oscillators.
However, from the structure of the matrix one sees that the operation of swapping the occupation amplitudes of the states $\ket{0}$ and $\ket{1}$ of qubit $a$ only, involves
subjecting the two pairs of oscillators $\ket{00},\ket{10}$ and $\ket{01},\ket{11}$ separately to the $\mathrm{SWAP}$ operation. This can be achieved by interaction for a time corresponding to a $\pi/2$ phase shift in the $\mathrm{SWAP}$ operation.

We will not consider the further transformations of Appendix \ref{sec:decompCNOT} explicitly since it is clear they can be performed analogously to the steps above. In this way all transformations of the $\mathrm{CNOT}$ quantum gate can be simulated by the classical oscillators. In any case, one sees that the complete $\mathrm{CNOT}$ gate
 \begin{equation}
 \mathrm{CNOT} = 
\begin{bmatrix}
1&0&0&0\\
0&1&0&0\\
0&0&0&1\\
0&0&1&0
\end{bmatrix}
\end{equation}
involves only a $\mathrm{SWAP}$ gate operation between the states $\ket{10}$ and $\ket{11}$ and so could be performed directly with classical oscillators. This is the advantage of having one directly-addressable oscillator for each quantum \emph{state}. The great disadvantage of course is that for $N$ qubits with two states each one needs a total of $2^N$ oscillators to perform the quantum simulation. This is one key element in the superiority of a quantum system in executing certain computing algorithms \cite{JoLi03_2011_}.

\section{The Landau-Zener Problem}
The aim of this section is to study time-dependent Hamiltonians. The Landau-Zener (LZ) problem of transitions between two  coupled quantum levels of varying energy is ubiquitous in quantum physics. LZ systems are characterised by the adiabatic i.e. time-independent eigenenergies of the coupled system exhibiting a typical avoided-crossing behaviour. 
Hence, first we show how this behaviour of the eigenenergies can be reproduced exactly by the eigenfrequencies of a pair of classical oscillators. Then we examine the time-dependent transition probability between the two coupled states and show that, to a very good approximation, the LZ transition also can be demonstrated with coupled oscillators.

\subsection{The Landau-Zener eigenenergies -- the general two level problem}
In section IIIB we considered the eigenvalues of a pair of  degenerate quantum levels interacting via an off-diagonal element $V$ and showed that the quantum eigenvalues $E$ are identical to the eigenfrequencies $\Omega$ of a pair of equal-frequency coupled oscillators, when we put $\hbar=1$. The simplest version of the Landau-Zener quantum problem involves two levels of varying energy, $E_1$ and  $E_2$, interacting via a fixed matrix element $V$. As the relative energy is varied the eigenvalues of the coupled system show an avoided crossing around $E_1=E_2$.  As a first step we show for the time independent case how to construct  a pair of classical oscillators whose eigenfrequency behaviour is identical to the quantum case.

The quantum coupled equations  (\ref{eq:Shcoupled}) have eigenvalues obtained by diagonalising the Hamiltonian matrix
 \begin{equation}
 \label{eq:QHam}
\mathbf{H} =  \left( \begin{array}{ c c }
E_1 & V \\	
V & E_2
 \end{array} \right),
 \end{equation}
 with the well-known result
 \begin{equation}
 \label{eq:Eval1}
E_{\pm} = \frac{1}{2}\left(E_1 + E_2 \pm \left[(E_1-E_2)^2 + 4V^2\right]^{1/2}\right).
\end{equation}
exhibiting the avoided crossing when $E_1=E_2$.
The classical Hamiltonian corresponding to this quantum Hamiltonian is
\begin{equation}
\label{eq:classHam3}
\mathcal{H} = \frac{1}{2}E_1(p_1^2 + q_1^2) + \frac{1}{2}E_2(p_2^2 + q_2^2) + Vq_1q_2 + Vp_1p_2.
\end{equation}
The Hamilton equations of motion are
\begin{eqnarray}
\begin{split}
&\dot q_1 = \omega_1p_1 + Vp_2, \qquad \qquad \dot q_2 = \omega_1p_2 + Vp_1\\
&\dot p_1 = -\omega_1 q_1 -  Vq_2,\qquad \dot p_2 = -\omega_2 q_2 -  Vq_1.
\end{split}
\end{eqnarray}
where we set $E_n = \omega_n$.

The Newton coupled equations resulting are
\begin{equation}
\label{eq:couplEqsLZ}
\begin{split}
\ddot q_1& + (\omega_1^2 + V^2)q_1 + V(\omega_1 + \omega_2)q_2 = 0\\
\ddot q_2 &+ (\omega_2^2 + V^2)q_2  + V(\omega_1 + \omega_2)q_1 = 0.
\end{split}
\end{equation}
 The eigenfrequencies are those of the matrix
 \begin{equation}
\mathbf{H^2} =  \left( \begin{array}{ c c }
E_1^2+V^2 & V(E_1+E_2) \\	
V(E_1+E_2) & E_2^2+V^2
 \end{array} \right),
 \end{equation}
 which are the eigenfrequencies $\Omega_{\pm}$ with, 
 \begin{equation}
 \label{eq:ExactEvals}
 \begin{split}
\Omega_{\pm}^2 = &\frac{1}{ 2}(E_1^2 + E_2^2+2V^2 \\
&\pm \left[(E_1^2-E_2^2)^2 + 4V^2(E_1+E_2)^2\right]^{1/2}).
\end{split}
\end{equation}
Although not immediately obvious it is readily shown that $\Omega_{\pm} = E_{\pm}$ of Eq.~(\ref{eq:Eval1}) as should be, since the eigenvalues of $\mathbf{H^2}$ are clearly the square of the eigenvalues of $\mathbf{H}$.

 As shown in the Appendix \ref{sec:standardClassicalEquations}, the standard equations describing coupled harmonic mass oscillators, when transformed to dimensionless coordinates, can be brought to the form
 \begin{equation}
\begin{split}
\label{eq:standard_oszi_X}
\ddot {X}_1& + \omega_1^2 X_1 - K\omega X_ 2 = 0\\
\ddot {X}_2& + \omega_2^2 X_2- K\omega X_1 = 0,
\end{split}
\end{equation}
where $\omega \equiv (\omega_1\omega_2)^{1/2}$.
By suitable choice of frequencies and couplings these equations can be put in the form of the exact mapping Eqs.~(\ref{eq:couplEqsLZ}) and hence the quantum eigenenergies can be simulated easily.

However, a more usual form of the standard equations for unequal frequency oscillators, derivable directly from the Hamiltonian Eq.~(\ref{eq:classHam3}) when the coupling term
$p_1p_2$ is neglected, is 
\begin{equation}
\label{eq:phieqns}
\begin{split}
\ddot{q}_1 + \omega_1^2 q_1& - K\omega_1 q_2 = 0\\
\ddot{q}_2 + \omega_2^2 q_2& - K\omega_2 q_1 = 0.
\end{split}
\end{equation}
It is clear that these equations are not identical to the exact mapping Eqs.~(\ref{eq:couplEqsLZ}) or to the Eqs.~(\ref{eq:standard_oszi_X}). The approximation involving the neglect of the momentum coupling and leading to these RCA equations is analysed in Paper 1. 
The RCA requires the validity of two approximations. The first is to assume a weak classical coupling such that $V/\omega_n \ll 1$ for $n=1,2$.  The second is to replace $V(\omega_1 + \omega_2)$ by $2V\omega_1$  or $2V\omega_2$ respectively in the off-diagonal coupling terms in Eqs.~(\ref{eq:couplEqsLZ}). This is valid when $\omega_1 \approx \omega_2$. Then the  RCA in Eqs.~(\ref{eq:couplEqsLZ}) gives the Newton equations (\ref{eq:phieqns}) with $K= -2V$. Replacing $E_1 \equiv \omega_1$ and $E_2\equiv \omega_2$ the eigenvalues of Eq.~(\ref{eq:phieqns}) are given by
 \begin{equation}
 \begin{split}
\Omega_{\pm}^2 = &\frac{1}{ 2}(E_1^2 + E_2^2 \\
&\pm \left[(E_1^2-E_2^2)^2 + 16V^2E_1E_2\right]^{1/2}).
\end{split}
\end{equation}
These eigenvalues are not equal to the exact ones of Eq.~(\ref{eq:ExactEvals}). However, when $E_1 = E_2 = E$ the exact eigenvalues are $\Omega_{\pm}^2 = (E\pm V)^2$ and the RCA values are $\Omega_{\pm}^2 = (E^2\pm 2VE)^2$, as shown above. 
The RCA expresses the approximation that $V \ll E$ so that in RCA $\Omega_{\pm} \approx (E\pm V)$, the exact values. Hence, when the de-tuning is not large i.e. $E_1 \approx E_2$,  it is possible to use classical oscillators satisfying the RCA "normal" equations (\ref{eq:phieqns}) rather than the exact equations (\ref{eq:couplEqsLZ}) to achieve a classical analogue of the quantum equations. The RCA equations involving only position couplings, are easy to realise practically e.g. with coupled masses or LCR circuits, which explains why most previous simulations of quantum systems have simply assumed this form.

In summary, when the two oscillators have different frequency as in the LZ problem, it is possible to construct a set of classical oscillators whose eigenfrequencies give the exact eigenvalues of the quantum LZ problem. The RCA  equations also will provide a reasonable approximation so long as the frequency difference is not great. Of course this is the case close to the avoided crossing in the LZ  coupled equations. As already mentioned, by comparing the Hamiltonians of Eq.~(\ref{eq:classHam2}) and Eq.~(\ref{eq:classHam3}) the RCA is equivalent to neglecting the off-diagonal momentum terms in the "quantum" Hamiltonian. \\

\subsection{The Landau-Zener Transition}

The classical/quantum mapping changes somewhat drastically when time-dependent Hamiltonians are admitted. Then upon differentiating the Schr\"odinger equation
\begin{equation}
\label{eq:Shcoupled_time}
i\mathbf{\dot{c}(t)} = \mathbf{H(t)}\mathbf{c(t)}
\end{equation}
with respect to time one has not Eq.~(\ref{eq:Clcoupled}) but the more complicated equation 
\begin{equation}
\label{eq:ClTDcoupled}
\mathbf{\ddot c} +\mathbf{H}^2\mathbf{c} + i\mathbf{\dot H}\mathbf{c} = 0
\end{equation}
The classical analogue is then written in the form
\begin{equation}
\label{eq:ddotzTD}
\mathbf{\ddot z}+ \mathbf{H}^2\mathbf{z} + i\mathbf{\dot H}\mathbf{z} = 0
\end{equation}
Using the Hamilton equations (\ref{eq:Hameqns}) one obtains the coupled Newton equations for the real displacements,
\begin{equation}
\label{eq:ClassLZ}
\mathbf{\ddot q} + \mathbf{H^2}\mathbf{q} + \mathbf{\dot H}\mathbf{H^{-1}}\mathbf{\dot q} = 0.
\end{equation}
Hence one sees that the time dependence of the quantum Hamiltonian has introduced new "forces" into the effective
classical equations of motion that involve the velocities and therefore appear as generalised frictional forces (although they do not have to be dissipative).\\

The standard two-level Hamiltonian corresponding to the LZ problem has the diagonal energies now time-dependent although the coupling $V$ is still constant. This gives
 \begin{equation}
 \label{eq:QHamTD}
\mathbf{H} =  \left( \begin{array}{ c c }
E_1(t) & V \\	
V & E_2(t)
 \end{array} \right),
 \end{equation}
 The time-dependent LZ problem consists of beginning
 in one state and calculating the probability amplitude for populating the other state as the avoided crossing is traversed.
 The classical equations of motion corresponding to the quantum LZ problem are obtained by substituting the Hamiltonian (\ref{eq:QHamTD}) in Eq.~(\ref{eq:ClassLZ}). In terms of the components $q_n$ this gives, with $E_n= \omega_n$
\begin{equation}
\label{eq:couplEqsLZTD}
\begin{split}
\ddot q_1 &+ (\omega_1^2 + V^2)q_1 + \frac{\dot\omega_1\omega_2}{(\omega_1\omega_2 - V^2)}\dot q_1 \\ 
& + V(\omega_2 + \omega_1)q_2 - \frac{\dot\omega_1V}{(\omega_1\omega_2 - V^2)}\dot q_2 = 0.\\
\ddot q_2 &+ (\omega_2^2 + V^2)q_2 + \frac{\dot\omega_2\omega_1}{(\omega_1\omega_2 - V^2)}\dot q_2 \\ 
& + V(\omega_1 + \omega_2)q_1 - \frac{\dot\omega_2V}{(\omega_1\omega_2 - V^2)}\dot q_1 = 0.
\end{split}
\end{equation} 
Compared to Eqs.~(\ref{eq:couplEqsLZ}) these equations have acquired  new diagonal and off-diagonal velocity coupling terms. Although in principle possible, it remains a challenge to find a real physical oscillator system with couplings that reproduce the quantum conditions. Note that the diagonal velocity coupling terms  can be removed from the equations  by a simple phase transformation but the off-diagonal velocity coupling cannot. Interestingly, however, in the RCA this direct velocity coupling term disappears also.

In RCA the Eqs.~(\ref{eq:couplEqsLZTD}) reduce to 
\begin{equation}
\label{eq:LZTDRCA}
\begin{split}
&\ddot q_1 + \omega_1^2 q_1 + \frac{\dot\omega_1}{\omega_1}\dot q_1+2 V\omega_1q_2 = 0\\
&\ddot q_2 + \omega_2^2 q_2 + \frac{\dot\omega_2}{\omega_2}\dot q_2 + 2V\omega_2q_1= 0
\end{split}
\end{equation} 
 Now the removal of the diagonal velocity terms by a phase transformation brings the equations to the standard q-coupled form realisable by linearly coupled oscillators. Hence it is interesting to test the validity of the RCA in the dynamic LZ problem of the traversal of the avoided crossing (see the next subsection). The RCA is only valid when $\omega_1 \approx \omega_2 \equiv \omega$ and $V \ll \omega$. Although the latter condition is easily satisfied, the former does not hold in general. Nevertheless the condition is valid precisely near the avoided crossing which is the region where the transition takes place. 
 
 \subsection{The Landau Zener Transition Probability}

 \begin{figure*}
\includegraphics[width=16cm]{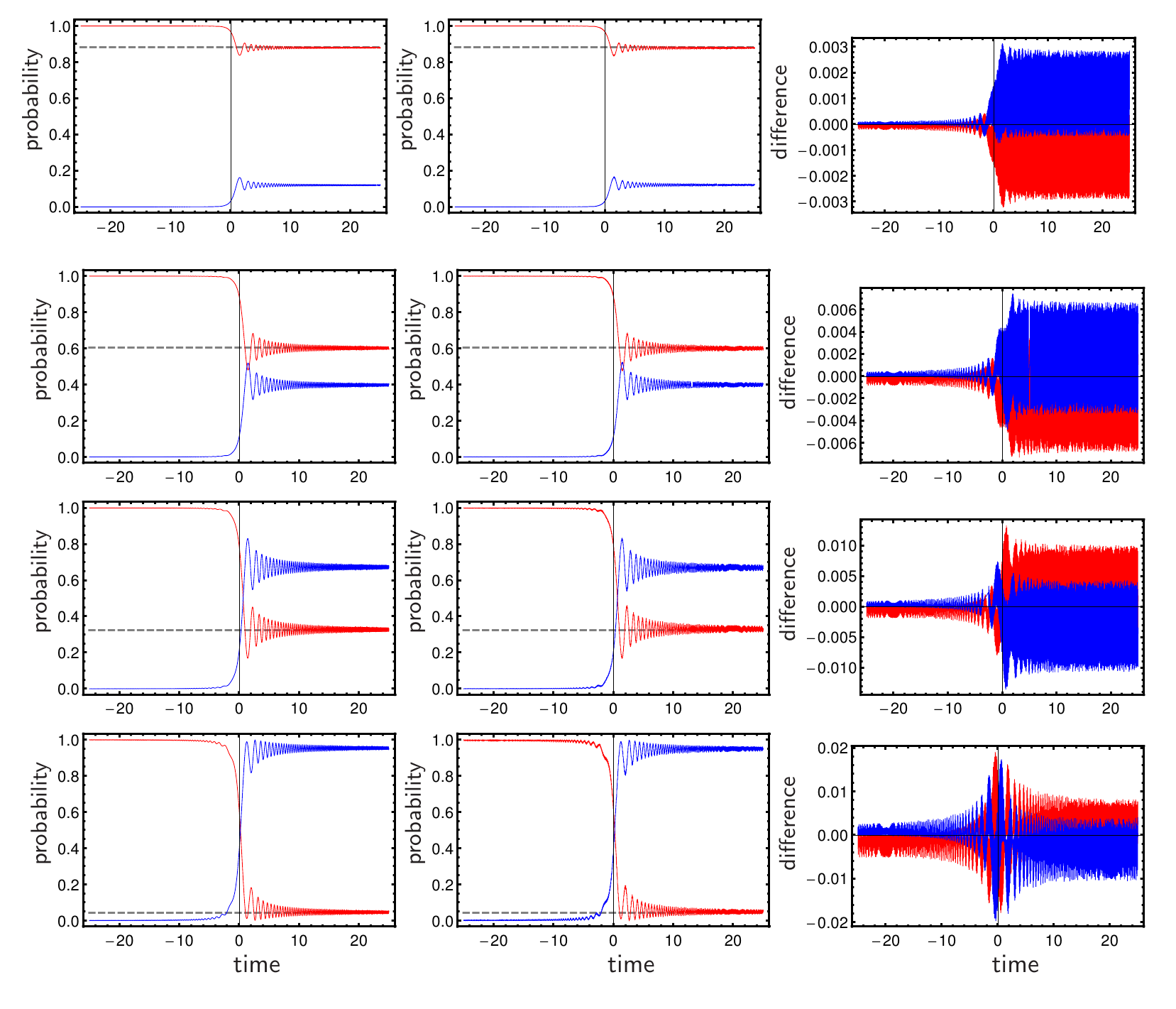}
\caption{\label{fig:FigLZ_RCA_Linear}
Occupation probability of state one (red) and two (blue).
Left column: Quantum result, Middle column: RCA, right column: difference between QM and RCA.
In all cases $E_{1/2}(t)=E_0\pm A \cdot t$. Time is given in units of $ \sqrt{\hbar/A} $ and energies are in units of $\sqrt{\hbar A}$. The initial time is taken as $t_0=-25$ and the energy $E_0=40$.
From top to bottom: $V= 0.2,\ 0.4,\ 0.6,\ 1.0$.
}
\end{figure*}

We consider the problem originally solved by Zener and St\"uckelberg \cite{Ze32_696_,St32_24_} where the time dependence of the crossing states is considered linear. That is, we take $E_1 = E_0 + At$ and $E_2 = E_0 - At$, where $E_0$ and $A$ are constants. The solution of the quantum LZ problem does not depend upon $E_0$ but for the RCA approximation to the classical equations to be valid we need to have $E_0 \gg V$ so that  we take $E_0$ as finite. Beginning at infinite negative time in state $1$, Zener's solution for the probability $P_2$ to occupy state $2$ at infinitely large positive times can be derived analytically as 
 \begin{equation}
\label{eq:PLZ}
P_2 = \exp{(-\pi V^2/A)}
\end{equation}
In Fig.~\ref{fig:FigLZ_RCA_Linear}  we compare $P_1(t) = |z_1(t)|^2$ and  $ P_2(t) = |z_1(t)|^2$ obtained from solving the RCA equations (\ref{eq:LZTDRCA}) with the same quantities obtained from the exact numerical solution of Eqs.~(\ref{eq:couplEqsLZTD}), which is of course the quantum solution. We show the exact and RCA results and explicitly the difference between them. For weak coupling $V$ (top row) the agreement of the RCA calculations with the exact quantum calculation is excellent, the difference being of the order of one half of a percent.
As the coupling becomes stronger however  the agreement is less good but never exceeds a difference of a few percent, even in the extreme case, where complete reversal of probabilities occurs (bottom row). 
The  choice of linear diabatic energies for all $t$, somewhat unphysical but necessary for the Zener analytic solution, actually turns out to be not ideally suited to simulation since, unless one chooses $E_0$ to be extremely large the possibility exists that the frequency of the classical oscillator can become negative. To remedy this, 
 we have also investigated a different time dependence that does not possess this unphysical behaviour by choosing the smooth function  $E_{1/2}(t)= 2 E_0 (1 \pm \arctan (t/E_0))$ which also shows a linear behaviour in the crossing region. Generally this leads to an even closer agreement between RCA and exact results.\\
 Several authors (e.g.~\cite{Jo98_397_,KoMaMa10_1281_,KoMaKo11_026602_}) have pointed out the similarity of quantum LZ equations to weakly-coupled classical oscillators involving only position couplings. Here we have shown that this correspondence requires the validity of the RCA.

\section{Dissipative states}
\label{sec:dissipative}
 There is an extensive literature on non-Hermitian Hamiltonians and the related questions of exceptional points in connection with the eigenvalue spectrum and the representation of environment coupling (for a recent review see Ref.~\cite{Ro13_178_}). A detailed discussion of this literature is beyond the scope of this article; suffice it to say that most aspects of this quantum physics can be simulated by classical oscillators. The connection of non-Hermitian quantum Hamiltonians to classical Hamiltonians has been discussed in detail by Graefe et.al.~\cite{GrHoeKo10_075306_}. The simulation of non-Hermitian and complex Hermitian quantum systems by classical oscillators is treated in a general way by Skinner \cite{Sk13_012110_}. Below we consider a simple example in detail. Basically, a quantum complex Hamiltonian operator has a corresponding complex classical Hamiltonian. Nevertheless as shown in Appendix \ref{sec:Non-hermitian-driven} one does not need to consider a Hamiltonian form since the TDSE leads directly to real Newton equations for the variables $\vek q$ and $\vek p$. 

Note that the approch presented in the present paper differs from the one that we used in Ref.~\cite{EiBr12_046118_} to treat open quantum systems, which was based on a stochastic unraveling of the reduced systems dynamic and results in the averaging over the dynamics oscillators with time dependent frequencies, dampings and couplings.

\begin{figure}[tp]
\includegraphics[width=8.cm]{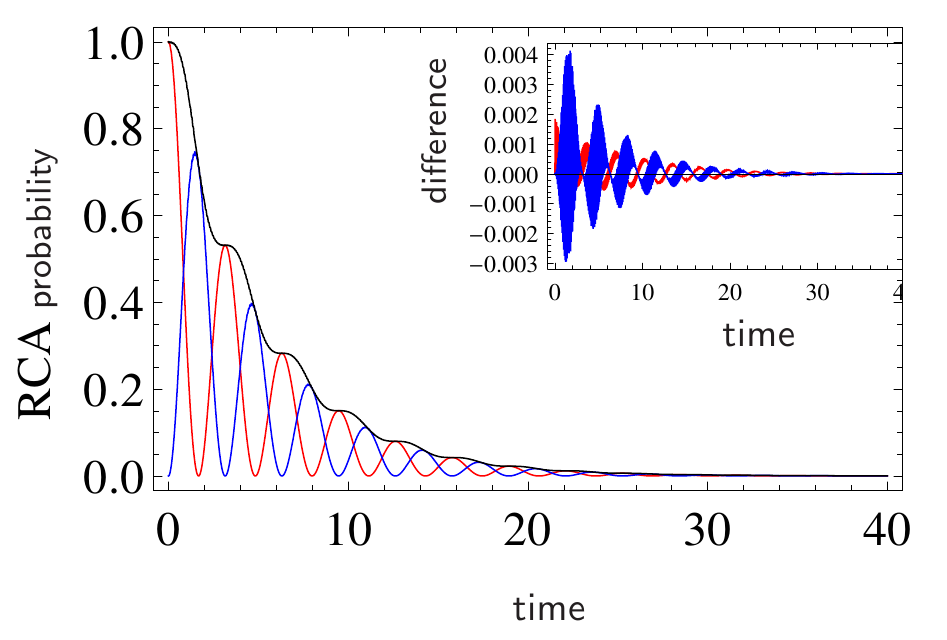}
\caption{\label{fig:damped-two-level}Coupled, damped two level system. 
The parameters are $E_1=E_2=40$, $\lambda_1=-0.0$, $\lambda_2=-0.2$, $V=1.0$.  
Red: $q_1^2+p_1^2$, Blue $q_2^2+p_2^2$, Black $(q_1^2+p_1^2)+(q_2^2+p_2^2)$.
The inset shows the difference between RCA and the exact quantum calculation.
}
\end{figure}

A common way of representing coupling to the environment in quantum mechanics is to make eigenergies complex leading to an effective damping term in the classical equations as shown also in Appendix \ref{sec:Non-hermitian-driven}. Hence,
as the simplest model of a dissipative quantum or classical system we take a two-state system having quantum Hamiltonian of the form
\begin{equation}
 \label{eq:QHamComplex}
\mathbf{H} =  \left( \begin{array}{ c c }
E_1 + i\lambda_1 & V \\	
V & E_2 + i\lambda_2
 \end{array} \right),
 \end{equation}
Despite its rather innocuous form this extension to complex Hamiltonian gives rise to coupled oscillator equations which, as in the time-dependent case, involve off-diagonal velocity couplings. The equation for $q_1$, derived in Appendix C is
\begin{equation}
\label{eq:Newt-nonherm}
\ddot q_1 + a\dot q_1 + b q_1 + c q_2 + d \dot q_2 =  0
\end{equation}
where the co-efficients are of the form (with $E \equiv \omega)$,
\begin{equation}
a \equiv -\left(\lambda_1 + \frac{(\omega_1\omega_2\lambda_1-V^2\lambda_2)}{(\omega_1\omega_2-V^2)}\right)
\end{equation}
\begin{equation}
b \equiv( \omega_1^2 + V^2) + \lambda_1\frac{(\omega_1\omega_2\lambda_1- V^2\lambda_2)}{(\omega_1\omega_2-V^2)}
\end{equation}
\begin{equation}
c \equiv V(\omega_1 + \omega_2) - \frac{(V\omega_1\lambda_1\lambda_2 -V\omega_1\lambda_2^2)}{(\omega_1\omega_2-V^2)}
\end{equation}
\begin{equation}
d \equiv  \frac{\omega_1V (\lambda_1 - \lambda_2 )}{(\omega_1\omega_2-V^2)}.
\end{equation}
 One readily sees that for no dissipation $\lambda_1= \lambda_2=0$ the equations reduce to the exact mapping 
equations Eqs.~(\ref{eq:couplEqsLZ}).\\ 
The momenta are given by (see Appendix C)
\begin{equation}
\label{eq:Newtmomenta}
\begin{split}
p_1 &= \frac{1}{(\omega_1\omega_2 - V^2)}~[ \omega_2\dot q_1 - V\dot q_2 - \lambda_1\omega_2q_1 + V\lambda_2 q_2]\\
p_2 &= \frac{1}{(\omega_1\omega_2 - V^2)}~[ \omega_1\dot q_2 - V\dot q_1 - \lambda_2 \omega_1q_2 +V\lambda_1q_1]
\end{split}
\end{equation}
More importantly, in the RCA, valid when $V \ll \omega_1,\omega_2$, which can always be realised for classical oscillators,
the equations reduce to the much simpler form,
\begin{equation}
\label{eq:couplEqsDiss}
\begin{split}
\ddot q_1& - 2\lambda_1\dot q_1 + (\omega_1^2 + \lambda_1^2)q_1 + V(\omega_1 + \omega_2)q_2 = 0\\
\ddot q_2 &- 2\lambda_2\dot q_2 +  (\omega_2^2 + \lambda_2^2)q_2  + V(\omega_1 + \omega_2)q_1 = 0.
\end{split}
\end{equation}
which are just the equations (\ref{eq:couplEqsLZ}) again but now with dissipation included. Similarly the momenta become
\begin{equation}
\begin{split}
p_1 &= \frac{(\dot q_1 - \lambda_1 q_1)}{\omega_1}\\
p_2 &= \frac{(\dot q_2 - \lambda_1 q_2)}{\omega_2}
\end{split}
\end{equation}
 Hence, as in the LZ case one can anticipate that these RCA equations give
an excellent reproduction of the quantum behaviour.  That this is indeed the case is shown in Fig.~\ref{fig:damped-two-level}  where we present the RCA result and the difference from the exact classical (and hence quantum)
result for exemplary realistic values of the dynamical parameters. The RCA is in error by less than one percent at all times.

\begin{figure}
\includegraphics[width=8.5cm]{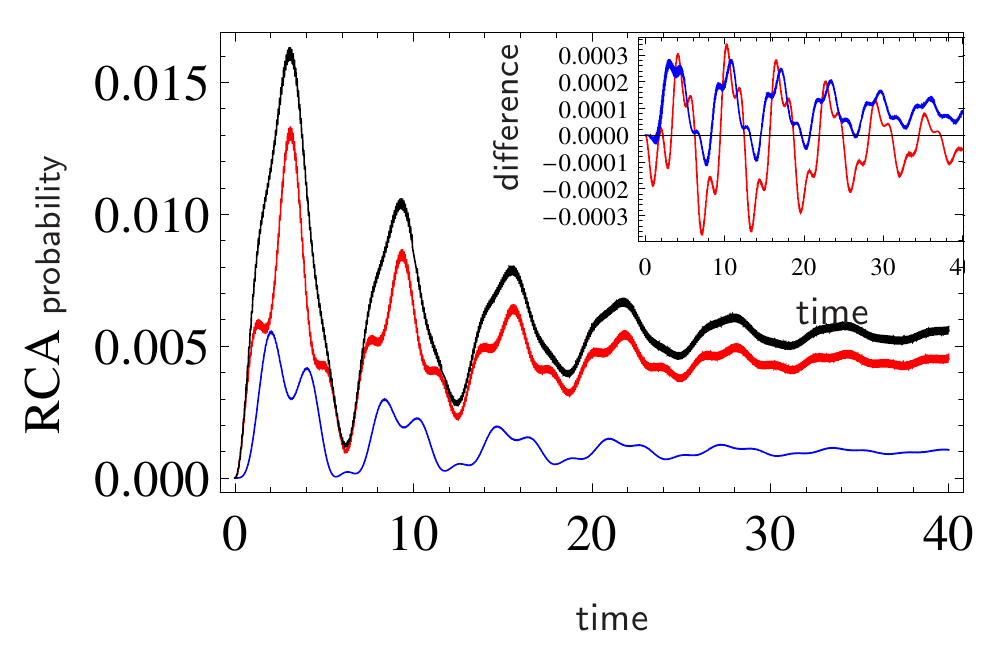}
\caption{\label{fig:damped-two-level-driven}
Coupled, damped, driven two level system. 
The parameters are the same as Fig \ref{fig:damped-two-level}. The driving strength of level $1$ is $\mu_1=0.2$ and initially both states have no population.
The inset shows the difference between RCA and the exact quantum calculation.}
\end{figure}

When subjected to external oscillatory forces it has been shown  that the response of a pair of coupled oscillators obeying the RCA equations Eq.~(\ref{eq:couplEqsDiss}) can simulate the profiles of typical "Fano"  interference resonances \cite{JoSaKi06_259_} and the phenomenon of EIT \cite{TaZhKo09_053901_}. From Appendix C one sees that the 
inclusion of an harmonic driving term with external frequency $\omega$ results in an inhomogeneous equation with the r.h.s. of Eq.~(\ref{eq:Newt-nonherm}) and  Eq.~(\ref{eq:couplEqsDiss}) simply replaced by appropriate 
terms $-\mu_{1/2}(\omega_{1/2} + V) \cos{\omega t}$, where $\mu_1$ and $\mu_2$ are the driving strengths from some initial state to levels 1 and 2 respectively . In Fig.~\ref{fig:damped-two-level-driven} we show again that 
the RCA gives excellent agreement with the quantum result for the occupation amplitudes of the two states as a function of time. Of course, when driven the amplitudes settle down to some steady state values.
The RCA error in the asymptotic values is of the order of  two percent. Here we have chosen the example of one oscillator, the driven one, having no damping and interacting with a damped second oscillator. This choice is similar to the one made to simulate the quantum situation of a Fano resonance where a narrow discrete state,
driven by the electromagnetic field, interacts with a broader continuum state. Were we to plot the response (asymptotic values in time) as a function of driving frequency we would obtain for appropriate parameters the typical Fano resonance profile. Again we note that previous simulations of Fano resonance and EIT phenomena have implicitly assumed the validity of RCA.

In the mapping of Dirac used here the real and imaginary parts of the quantum amplitude $c_n(t)$ are identified with the $q$ and $p$ variables respectively
 of a \emph{single} classical oscillator. Skinner \cite{Sk13_012110_} has made the suggestion that, since both $q$ and $p$ vary harmonically, one can use each as an independent oscillator variable. This has the clear disadvantage that the number of  classical oscillators required is doubled (e.g. in the $\mathrm{SWAP}$ quantum gate one would require eight coupled oscillators). However, in certain cases it can be advantageous. In particular let us map the above Hamiltonian of Eq.~(\ref{eq:QHamComplex}) in this way. The coupling of two quantum states now is simulated by four oscillators $ \vek q = q_1,q_2$ and $\vek p = p_1,p_2$. The coupled Newton equations read now (we present only $q_1$  and $p_1$, the other two equations are analogous),
 \begin{equation}
\begin{split}
\ddot q_1 + (\omega_1^2+V^2-\lambda_1^2)q_1 + &V(\omega_1+\omega_2)q_2 - 2\omega_1\lambda_1p_1\\& - V(\lambda_1+\lambda_2)p_2 = 0\\
\ddot p_1 + (\omega_1^2+V^2-\lambda_1^2)p_1 + &V(\omega_1+\omega_2)p_2 + 2\omega_1\lambda_1q_1\\& + V(\lambda_1+\lambda_2)q_2 = 0.
\end{split}
\end{equation}
Although four oscillators are involved, these equations probably are simpler to realise with actual oscillators than the two coupled oscillators described by Eq.~(\ref{eq:Newt-nonherm}) which involve off-diagonal velocity coupling.\\

\section{Conclusions}
The quantum dynamics of the complex amplitudes $c_n(t)$ of $N$ coupled quantum states can be mapped, via $c_n(t) \equiv( q_n(t) + ip_n(t))/\sqrt 2$, onto the classical dynamics of $N$ coupled oscillators.  This result
is completely independent of the character (e.g. single-particle or many-particle) nature of the quantum states. 
The equivalent classical Hamiltonian is a function of the quantum Hamiltonian matrix elements in the basis of $N$ eigenstates. Depending upon the nature of the matrix elements
the requirements for the simulation by realisable classical oscillator systems can be straightforward or more difficult. For real hermitian quantum Hamiltonians the simulation \emph{is}
straightforward and we have shown explicitly how, using the molecular J aggregate as example, entanglement measures based on pure states can be simulated by classical oscillators. Further we showed that classical qubits can be defined and Bloch rotations and all fundamental two qubit gate operations performed also by coupling classical oscillators. However, since from $N$ qubits one can build a total of $2^N$ quantum states, then, as $N$ itself becomes large, one would need an exceedingly large number  of $2^N$ oscillators to achieve the simulation. Of course, this is precisely the departure point between quantum and classical mechanics.
Also, if the quantum system is composed of individual subsystems a corresponding decomposition of the system of classical oscillators can usually not be achieved. That is, in general there is no one-to-one correspondence between classical oscillator and quantum subsystem.

One should also mention the difference in the measurement process required to ascertain the complex amplitudes. In the classical case, since the dynamics is deterministic, the amplitude and phase can be measured directly. In the quantum case the same numbers are obtained as the statistical averages of many measurements. For example in the case of the SQiSW quantum gate employed in \cite{BiAnHo10_409_}, of the order of one thousand measurements are required to achieve the necessary accuracy.

We have considered the extension to time-dependent and non-hermitian quantum Hamiltonians, using the two-state LZ problem and decaying states as the simplest examples.
Again, here an exact mapping is possible but gives rise to more complicated Newton equations of motion involving off-diagonal velocity couplings. Nevertheless, in principle these terms can be simulated, either by damping or driving, as in the case of electrical oscillator circuits with negative resistance, for example. However, in the RCA, which requires that the coupling between oscillators is weak, the
 complicated exact equations simplify to a standard form without such off-diagonal velocity couplings. We have shown that the RCA equations give excellent agreement with exact quantum results when the weak coupling criterion is satisfied.

  Recently,  an
 alternative way to achieve standard coupled-oscillator classical equations has been proposed\cite{Sk13_012110_}. This is to recognise that the $p_n$ variables in the uncoupled limit are also harmonic and therefore to treat the momenta as the position variables of $N$ additional classical oscillators. Although this doubles the number of independent oscillators required for the mapping, it has the advantage that the troublesome velocity couplings are eliminated.  Hence, although not necessary for real hermitian Hamiltonians, doubling the number of oscillators can be advantageous for non-hermitian (and also for time-dependent)  Hamiltonians when the number of states $N$ is not large. Which of the two schemes is  simpler to realise in practice will depend upon the precise nature of the quantum problem at hand.
Without claiming to encompass all possible scenarios, we would maintain that in most cases of practical simulation, the use of $N$ coupled oscillators satisfying the RCA, leading to standard coupled equations as illustrated above will furnish sufficient accuracy. This is because weak coupling is necessary to maintain the linearity of the oscillator and coupling forces with displacement i.e. to satisfy Hooke's Law or its electrical equivalent. If higher accuracy is required one must additionally simulate the couplings neglected in the RCA or resort to $2N$ oscillators according to Skinner's prescription \cite{Sk13_012110_}. However, it should always be remembered that the truncated set of \emph{quantum} equations in the examples used here is also an approximation. Any actual quantum system would show deviations from the predictions of our Schr\"odinger equation since inevitable coupling to states not included in the truncated basis is not taken into account. This is the quantum analogue of non-linear terms neglected in the classical simulation and, as in that case, would be a more serious approximation when the coupling becomes strong.

\section{Acknowledgement}
We acknowledge very useful comments on an earlier version of the manuscript from Prof. T.E. Skinner. JSB is grateful to Prof.\ J-M. Rost and the Max-Planck-Institute for Physics of Complex Systems for their hospitality and to Prof. Hanspeter Helm for many helpful discussions.

\appendix

\section{Decomposition of the CNOT gate}
\label{sec:decompCNOT}
The decomposition of the CNOT two qubit quantum gate is
\begin{equation}
\begin{split}
R_y^a(-\pi/2)[ R_x^a(\pi/2) \otimes R_x^b(-\pi/2)] &\\\mathrm{SQiSW} R_x^a(\pi) \mathrm{SQiSW} R_y^a(\pi/2)
\end{split}
\end{equation}
We follow this sequence of transformations through and
 as example we
consider the state $\ket{00}$ as  the initial state. 
The normalised one qubit states are then initially
\begin{equation}
\ket{\psi_{a,b}} = 1 \ket{0}_{a,b} + 0 \ket{1}_{a,b}   
\end{equation}
and the two-qubit state the separable product
\begin{equation}
\label{eq:Psiab1}
\ket{\Psi_{ab}}_1 = \ket{\psi_a}\ket{\psi_b} =  1\ket{00} + 0\ket{01} + 0\ket{10} + 0\ket{11}.
\end{equation}
The first rotation $R_y^a(\pi/2)$  results in the excitation of state $\ket{10}$, i.e.
\begin{equation}
\label{eq:Psiab2}
\ket{\Psi_{ab}}_2 = \frac{1}{\sqrt 2}\ket{00} + 0\ket{01} + \frac{1}{\sqrt 2}\ket{10} + 0\ket{11}.
\end{equation}
The SQiSW operation entangles only the two $\ket{01}$ and $\ket{10}$ states and in this 2-dimensional space
 this entanglement operation does not affect the $\ket{00}$ and $\ket{11}$ states. There results the non-separable state
\begin{equation}
\label{eq:Psiab3}
\ket{\Psi_{ab}}_3 = \frac{1}{\sqrt 2}\ket{00} - \frac{i}{2}\ket{01} + \frac{1}{2}\ket{10} + 0\ket{11}.
\end{equation}
The interaction is now switched off and a further one-qubit rotation $R_x^a(\pi)$
performed on qubit $a$, to give
\begin{equation}
\label{eq:Psiab4}
\ket{\Psi_{ab}}_4 = -\frac{i}{2}\ket{00}  + 0\ket{01} - \frac{i}{\sqrt 2}\ket{10} - \frac{1}{2}\ket{11}.
\end{equation} 
The second SQiSW entanglement step gives,
\begin{equation}
\label{eq:Psiab5}
\ket{\Psi_{ab}}_5 = -\frac{i}{2}\ket{00} - \frac{1}{2}\ket{01} - \frac{i}{2}\ket{10} - \frac{1}{2}\ket{11}.
\end{equation}
 The two independent qubits are now simultaneously rotated by angles $\pi/2$ and $-\pi/2$ respectively about the $x$ axis. There results
 the state
 \begin{equation}
 \label{eq:Psiab6}
\ket{\Psi_{ab}}_6 = -\frac{1}{2}\left((1+i)\ket{00} + 0\ket{01}  + (1+i)\ket{10} + 0\ket{11}\right).
\end{equation}
The final rotation of qubit $a$ alone results in the initial state, to within a global phase factor, i.e.
\begin{equation}
\label{eq:Psi7}
\begin{split}
\ket{\Psi_{ab}}_7 &= -\frac{1+i}{\sqrt2}\ket{00} + 0\ket{01} + 0\ket{10} + 0\ket{11}\\
&= e^{(-i\pi/4)}( 1\ket{00} + 0\ket{01} + 0\ket{10} + 0\ket{11}).
\end{split}
\end{equation}  
\\
It is interesting to note that the second SQiSW operation is actually a dis-entangling step since $\ket{\Psi_{ab}}_5$ of Eq.(\ref{eq:Psiab5}) is the
separable state
\begin{equation}
\ket{\Psi_{ab}}_5 = \ket{\psi_a} \ket{\psi_b} = -\frac{i}{2} ( \ket{0}_a + \ket{1}_a)( \ket{0}_b -i \ket{1}_b)
\end{equation}
Therefore the subsequent  one qubit rotation operations can be performed separately on these  states. It is easy to check that for each of the four CNOT operations of Eq.(\ref{eq:CNOT}) after the second SQiSW a separable state is obtained. This must be so, since the final target states are separable and the  single qubit rotations subsequent to the second SQiSW cannot induce entanglement.\\

 \section{Standard classical equations}
\label{sec:standardClassicalEquations}
 Here we derive the standard classical equations of motion for two harmonic oscillators of masses $m_1$ and $m_2$ coupled by a spring. The Hamiltonian is taken as
 \begin{equation}
 \label{eq:unscalHam}
\mathcal{H} = \frac{{\tilde p_1}^2}{2m_1} + \frac{1}{2}m_1\omega_1^2{\tilde x}_1^2 + \frac{{\tilde p_2}^2}{2m_2} + \frac{1}{2}m_2\omega_2^2{\tilde x}_2^2 - \kappa \tilde x_1\tilde x_2
\end{equation} 
The scaling to dimensionless variables $(x,p)$ is achieved by the transformation
\begin{equation}
\label{eq:scaling}
\tilde x_n = (\frac{m_n\omega_n}{\hbar})^{1/2}x_n ~,\qquad \tilde p_n = \frac{1}{(m_n\hbar \omega_n)^{1/2}}p_n
\end{equation}
for $n=1,2$. This gives the new Hamiltonian
\begin{equation}
\label{eq:classHam2}
\mathcal{H}/\hbar = \frac{1}{2}\omega_1(p_1^2 + x_1^2) + \frac{1}{2}\omega_1(p_2^2 + x_2^2) - K x_1x_2
\end{equation}
where
\begin{equation}
K \equiv \frac{\kappa}{(m_1m_2\omega_1\omega_2)^{1/2}}
\end{equation}\\
and all terms are of the physical dimension of inverse time. Here we have included $\hbar$ explicitly so that the connection to the "quantum" Hamiltonian Eq.(\ref{eq:classHam3}) is obvious. With $E_n = \hbar\omega_n$ they are of the same form except that the classical Hamiltonian is missing the p-coupling terms. Hence we call this expression the q-coupled Hamiltonian. With this Hamiltonian the equations of motion are 
\begin{eqnarray}
\begin{split}
&\dot x_1 = \omega_1p_1, \qquad \qquad \dot x_2 = \omega_1p_2\\
&\dot p_1 = -\omega_1 x_1 +  Kx_2,\qquad \dot p_2 = -\omega_2 x_1 +  Kx_1.
\end{split}
\end{eqnarray}
From these equations are derived the coupled Newton equations
\begin{equation}
\begin{split}
\label{eq:NewtonRCA}
\ddot{x}_1 + \omega_1^2 x_1& = K\omega_1 x_2\\
\ddot{x}_2 + \omega_2^2 x_2& = K\omega_2 x_1.
\end{split}
\end{equation}
Note that we have chosen $\hbar$ in the scaling to make contact with the quantum Hamiltonian but since the final equations do not depend on it, we could have chosen any other constant with the dimension (Energy $\times$ Time) to fix the units.\\

Alternatively we can take the Hamiltonian  Eq.~(\ref{eq:unscalHam})  
and transform
\begin{equation}
\begin{split}
X_1& = \left (\frac{m_1}{m_2}\right)^{1/4}\tilde{x}_1, \qquad X_2 = \left (\frac{m_2}{m_1}\right)^{1/4}\tilde{x}_2\\
P_1 & = \left (\frac{m_2}{m_1}\right)^{1/4}\tilde{p}_1,\qquad P_2 =  \left (\frac{m_1}{m_2}\right)^{1/4}\tilde{p}_2
\end{split}
\end{equation}
This gives the completely symmetric form
\begin{equation}
\mathcal{H} = \frac{{ P_1}^2}{2\mu} + \frac{1}{2}\mu\omega_1^2X_1^2 + \frac{{P_2}^2}{2\mu} + \frac{1}{2}\mu\omega_2^2X_2^2 - \kappa X_1X_2
\end{equation}
with $\mu = (m_1m_2)^{1/2}$. The Newton equations are 
\begin{equation}
\begin{split}
\ddot {X}_1& + \omega_1^2 X_1 - K\omega X_ 2 = 0\\
\ddot {X}_2& + \omega_2^2 X_2- K\omega X_1 = 0,
\end{split}
\end{equation}
where we define a mean frequency $\omega \equiv (\omega_1\omega_2)^{1/2}$ so that $K = \kappa/(\mu\omega)$ as above.
 With this form one can satisfy the exact mapping equations Eqs.~(\ref{eq:couplEqsLZ}) since the coefficient of the coupling term is the same
in the two equations. If we scale the lengths $X_n$ to become dimensionless  by different factors for each $n$ as in Eqs.~(\ref{eq:scaling}), then we will achieve equations of the standard form Eqs.~(\ref{eq:NewtonRCA}). However, if we choose a common scaling factor for length e.g. $[\hbar/(\mu\omega)]^{1/2}$, then the equations are unchanged and in particular the coupling terms retain their common coefficient. This allows an exact mapping of the quantum Landau Zener equations for fixed energies.

\section{Non-hermitian driven two level system}
\label{sec:Non-hermitian-driven}

The Schr\"odinger equation is
\begin{equation}
\label{eq:Shcoupled2}
\mathbf{\dot{c}} =  -i\mathbf{H}\mathbf{c}.
\end{equation}
With  $\mathbf{H} = \mathbf{H_R} + i \mathbf{H_I}$  this  gives real first-order equations \cite{Sk13_012110_},
\begin{equation}
\label{eq:hameq2}
\mathbf{\dot q} = \mathbf{H_R}\mathbf{p} + \mathbf{H_I} \mathbf{q} \qquad \mathbf{\dot p} = -\mathbf{H_R} \mathbf{q} + \mathbf{H_I}\mathbf{p}
\end{equation}
The Newton equations for the real  oscillator amplitudes $q(t)$ are then
\begin{equation}
\mathbf{\ddot q} = -\mathbf{H_R}^2\mathbf{ q}  + \mathbf{H_I} \mathbf{\dot q} + \mathbf{H_R} \mathbf{H_I} \mathbf{H_R}^{-1} \mathbf{\dot q}  - \mathbf{H_R} \mathbf{H_I} \mathbf{H_R}^{-1} \mathbf{H_I}\mathbf{q}
\end{equation}
and the real momenta $p(t)$ are given by
\begin{equation}
\mathbf{p} = \mathbf{H_R}^{-1} (\mathbf{\dot q} -  \mathbf{H_I}\mathbf{ q}).
\end{equation}
The Hamiltonian matrices for a two-level system with $E \equiv \omega$ are
\begin{equation}
 \label{eq:QHamRe}
\mathbf{H_R} =  \left( \begin{array}{ c c }
\omega_1  & V \\	
V & \omega_2 
 \end{array} \right),
 \end{equation}
and
\begin{equation}
 \label{eq:QHamIm}
\mathbf{H_I} =  \left( \begin{array}{ c c }
\lambda_1 & 0 \\	
0 & \lambda_2
 \end{array} \right).
 \end{equation}
The inverse of $\mathbf{H_R}$ is 
\begin{equation}
 \label{eq:QHamInv}
\mathbf{H_R}^{-1} = \frac{1}{(\omega_1\omega_2 - V^2)} \left( \begin{array}{ c c }
\omega_2 & -V \\	
-V & \omega_1
 \end{array} \right).
 \end{equation}
With these definitions, the Newton equations become those of Eqs.~(\ref{eq:Newt-nonherm}) and Eqs.~(\ref{eq:Newtmomenta}) of section V.\\
When driven by an oscillating external field of frequency $\omega$, the Schr\"odinger equation becomes
\begin{equation}
\label{eq:ShDriven}
\mathbf{\dot{c}} =  -i\mathbf{H}\mathbf{c} -i \mathbf{f}(t),
\end{equation}
where 

\begin{equation}
\mathbf{f} =  \cos{\omega t}\left( \begin{array}{ c  }
\mu_1  \\	
 \mu_2
 \end{array} \right),
 \end{equation}
 and $\mu_1$ and $\mu_2$ are the driving strengths from some initial state. The Newton equations arising similarly have an additional inhomogeneous term equal to
 $- \mathbf{H_R}\mathbf{f}(t)$.


\end{document}